%% file: main.tex
\documentclass[sigconf]{acmart}
\setcopyright{none}
\usepackage{booktabs}      
\usepackage{multirow}      
\usepackage{balance}       
\usepackage{listings}      
\usepackage{xcolor}        
\usepackage{xspace}        
\usepackage{microtype}     
\usepackage{graphicx}      
\usepackage{subcaption}    
\usepackage{amsmath,amsfonts} 
\usepackage{enumitem}      
\usepackage{pifont}        
\usepackage{etoolbox}      

\input{macros}

\title{Toward a Causal Data Management Ecosystem for Decision Making and Agentic AI}

\author{Dazhuo Qiu}
\affiliation{%
  \institution{Lyon 1 University, CNRS Liris}
  \city{Lyon}
  \country{France}
}
\email{dazhuo.qiu@univ-lyon1.fr}

\author{Yingli Zhou}
\affiliation{%
  \institution{Lyon 1 University, CNRS Liris}
  \city{Lyon}
  \country{France}
}
\email{yingli.zhou@univ-lyon1.fr}

\author{Amedeo Pachera}
\affiliation{%
  \institution{Lyon1 University, CNRS Liris}
  \city{Lyon}
  \country{France}
}
\email{amedeo.pachera@univ-lyon1.fr}

\author{Angela Bonifati}
\affiliation{%
  \institution{Lyon1 University, CNRS Liris \& IUF}
  \city{Lyon}
  \country{France}
}
\email{angela.bonifati@univ-lyon1.fr}

\author{Andrea Mauri}
\affiliation{%
  \institution{Lyon1 University, CNRS Liris}
  \city{Lyon}
  \country{France}
}
\email{andrea.mauri@univ-lyon1.fr}

\begin{abstract}
Modern AI is no longer a single model but an ecosystem: classical ML predictors, deep and multimodal models, large language models, and agents, each trained and tuned over different data sources and each producing outputs at scale that become inputs to the others. Operating such an ecosystem is fundamentally a data integration problem—the knowledge it depends on is fragmented across dozens of heterogeneous, independently governed sources that must be reconciled and continually maintained. Yet integration alone is not enough. The predictions these systems make are shaped by many interacting factors, and the events, decisions, and variables that drive an outcome are routinely entangled with the ones that merely accompany it; treated as a basis for action, such correlational signals invite confounded decisions. This becomes acute once agents act autonomously: to be trustworthy and reliable, an agent must anticipate the consequences of its actions, not merely extrapolate from what has co-occurred before. Causal reasoning is what closes this gap, distinguishing the drivers of an outcome from its correlates, and enabling prescriptive and counterfactual analysis over the ecosystem's data. We therefore argue that the integrated ecosystem needs an explicit causal layer, and we propose to build it as a shared, persistent, queryable Causal World System (CWS).
\end{abstract}

\begin{document}

\maketitle

\section{Introduction}

A decade ago, deploying AI meant shipping a model. Today it means operating an \emph{ecosystem} where classical ML and deep-learning predictors, large language models, retrieval pipelines, and agents run side by side, each trained and tuned over many heterogeneous sources, such as tables, graphs, documents, images, audio streams, logs, support tickets, code repositories, and business applications—and each producing predictions and actions at scale that become inputs to the others. Keeping such an ecosystem coherent is, first of all, a \emph{data management} problem: the relevant knowledge is fragmented across dozens of independently governed sources that must be cleaned, aligned, integrated, and continually maintained. 

A modern enterprise or research lab already exposes its systems to all of these sources through a data layer, which makes the sources jointly queryable, joinable, and available as training and test data. 
Despite their success, current data management systems primarily exploit
correlations and statistical dependencies in the observed data, while causal
relationships, interventions, and counterfactuals are not represented as
first-class system primitives.
This gap matters because integration alone is not enough. 
Many interacting factors shape the outputs AI systems produce, but not all of these factors carry equal weight: some are true drivers of an outcome, while others merely accompany it without causing it.



While an application can observe that revenue fell after a price change, that a complaint spike followed a product update, or that a model regressed after fine-tuning, it cannot tell whether the change \emph{caused} the outcome or what would have happened otherwise. If not trained with an explicit notion of causality, models may conflate correlation with ~\cite{pearl2009causality,scholkopf2021crl}, and consider spurious shortcuts~\cite{li2024causaldl} in decision making processes.


This becomes critical when AI-based systems aren't used only for predicting and are deployed to \emph{act} autonomously. As agents change the very systems they observe, prediction is no longer sufficient: to be trustworthy and reliable, an agent must anticipate the consequences of its own actions—what its next action will set in motion, and what would have followed from a different one rather than extrapolate from what has co-occurred before~\cite{lecun2022path,nam2026cjepa}. We argue that causal reasoning is what closes this gap. By distinguishing the drivers of an outcome from its correlates, it supports prescriptive analytics and actionable predictions for humans, and counterfactual evaluation of candidate actions for agents. We therefore argue that the integrated ecosystem needs an explicit \emph{causal layer}.

\begin{figure*}[t]
\centering
\Description[overview]{}{}
\includegraphics[width=0.92\textwidth]{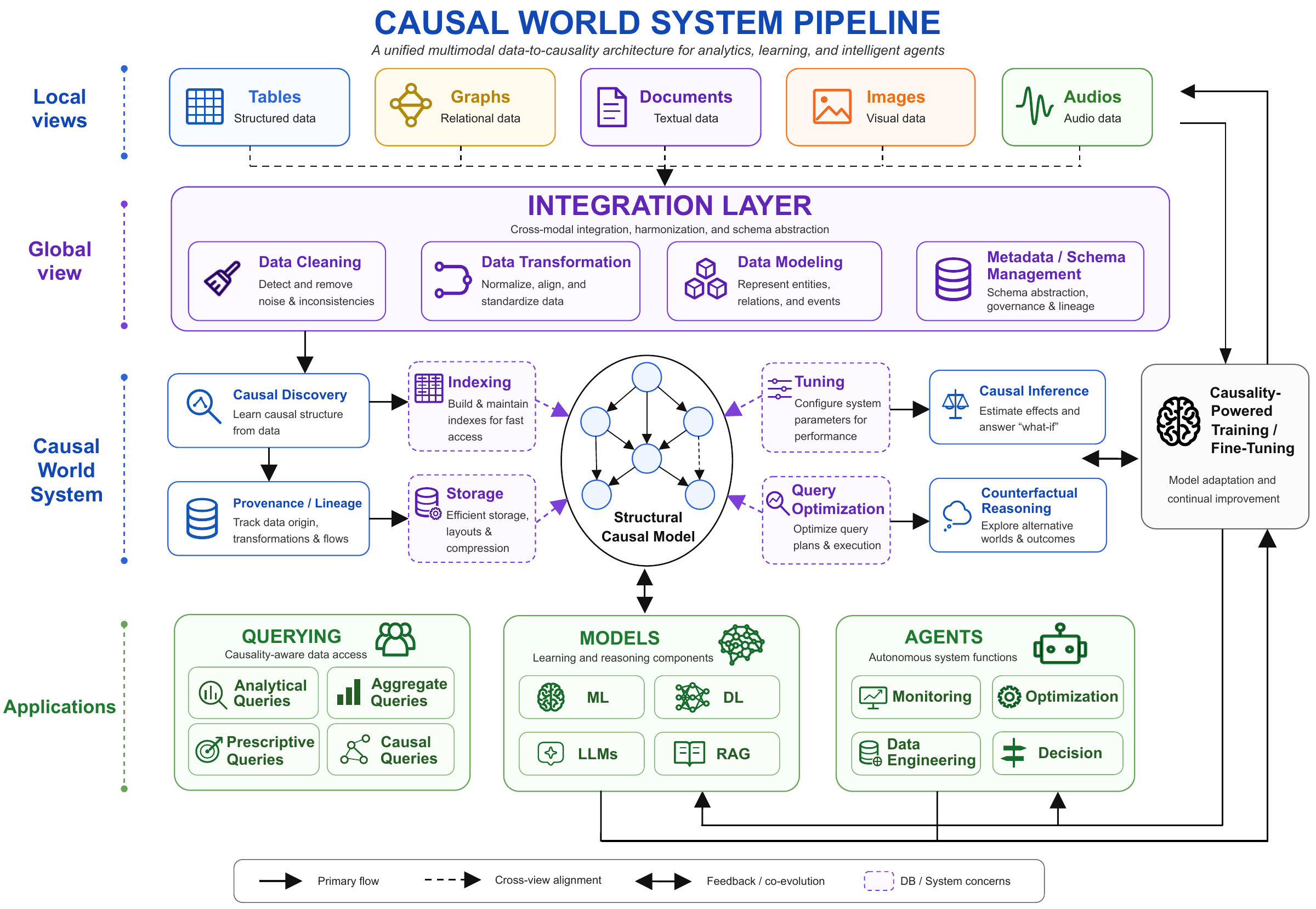}
\caption{Causal World System Overview}
\label{fig:arch}
\end{figure*}

Two observations shape how such a layer must be built. First, the causal knowledge it needs does not live in one modality or database. It is latent in the relationships \emph{between} local views (e.g. an abstraction and selection over the data): a campaign in a marketing platform, an order in a commerce database, a complaint in a support system, an image or document attached to a ticket, and a churn event in the CRM. Answering a causal question therefore requires cleaning, transforming, modeling, and aligning heterogeneous data sources before causal discovery, inference, and counterfactual reasoning can even begin. This makes causality for the ecosystem fundamentally a data management problem, apart from being a learning problem.

Second, causal knowledge must serve many consumers at once. Analysts need aggregate and prescriptive queries; machine-learning systems need structured causal signals for training and fine-tuning; agents need local estimates of the effects of candidate actions; and operators need ecosystem-level diagnoses. A single monolithic latent causal model is poorly suited to this range of needs. What is required instead is a persistent substrate that can expose local causal views, integrate them into a global causal structure, preserve provenance and lineage, and support interventional and counterfactual queries at multiple levels of abstraction.

This proposal builds on, but differs from, several related threads.
\emph{Causal machine learning} formalizes data generation as structural causal models and studies interventions and counterfactuals across supervised, generative, and reinforcement-learning settings~\cite{kaddour2022causalml,pearl2009causality}; 
recent surveys document how causal structure can improve robustness and interpretability over purely correlation-based deep learning~\cite{li2024causaldl,scholkopf2021crl}. We build on this foundation, but shift the unit of analysis from a single model or task to an entire AI and data ecosystem. \emph{World models} and JEPA-style architectures learn predictive latent dynamics for perception and control~\cite{lecun2022path,assran2023ijepa,bardes2023mcjepa,chen2025vljepa,nam2026cjepa};
we contrast these implicit, single-model substrates with an explicit, shared, queryable causal substrate. Finally, \emph{data integration} contributes the mediated-schema and answering-queries-using-views machinery that we lift into causality~\cite{lenzerini2002data,halevy2001views}, while data fusion~\cite{bareinboim2016fusion} and causal abstraction~\cite{beckers2019abstraction} supply tools for combining heterogeneous evidence and reasoning across levels. 
To our knowledge, no prior work proposes a mediated \emph{causal} schema as shared infrastructure for an ecosystem of agents, models, analysts, and decision-makers.

We therefore propose treating cause and effect as shared ecosystem \emph{infrastructure}. We call this model the \textbf{Causal World System (CWS)}: a persistent, explicit, queryable causal model that overlays the architectures, systems, and processes of an organization, a lab, or the broader AI ecosystem. As illustrated in Figure~\ref{fig:arch}, the CWS connects multimodal local views (e.g. heterogeneous data sources) through a data integration layer, constructs and maintains a causal DAG with provenance, and exposes this structure to querying systems, predictive models, and autonomous agents. The resulting feedback loop enables causality-powered training~\cite{kocaoglu2018causalgan,geiger2022NN}, fine-tuning~\cite{zhang2026causal}, monitoring~\cite{zeng2025DAGsumm}, optimization~\cite{mansouri2024object}, data engineering~\cite{yaramakala2005speculative,chen2024using}, and decision-making~\cite{pearl2009causality}.

\section{The Vision}
\label{sec:vision}

A Causal World System is the ecosystem's answer to the questions like \textit{Why} (explanation), \textit{What if I do} (intervention - updates), and \textit{What I would have done} (counterfactual). 

Concretely, it is composed by a structural causal model (SCM)~\cite{pearl2009causality} whose variables are the meaningful quantities of the organization—prices, inventory, latencies, ticket volumes, sentiment, churn, experiment arms—and whose mechanisms encode how interventions on some of these propagate to the rest. It is not a static analytics dashboard but an artifact that humans, agents, and learning systems query and interact with continuously. Across all three, causality is what upgrades each consumer: it turns description into prescription for humans, prediction into deliberation for agents, and correlation-fitting into structure-aware learning for the models themselves. Figure~\ref{fig:arch} sketches how this plays out for each consumer: humans through the querying interface, models through causality-powered training, and agents through counterfactual self-modeling.

\paragraph{Humans: across the full query spectrum.} 
Decision-makers want more than a forecast; they want to know which lever to pull. A CWS serves a graded spectrum of questions over the same substrate—from \emph{analytical} and \emph{aggregate} queries (how did churn evolve last quarter?''), to \emph{prescriptive} ones (which intervention minimizes churn under budget?''), to explicitly \emph{causal} and \emph{counterfactual} ones. Instead of support tickets will rise next week,'' it answers raising the price by $5\%$ will raise tickets by $12\%$ and churn by $3\%$, but discounting shipping offsets two-thirds of that.'' These are \emph{actionable} predictions, estimates of the effect of an intervention, ranked by the outcomes a human actually controls.

\paragraph{Agents: from prediction to counterfactual self-modeling.}
An agent embedded in the ecosystem can consult the CWS \emph{before} acting, simulating the interventional distribution of its candidate actions and comparing counterfactual outcomes. This converts an agent from a reactive predictor into a deliberative actor that can ask ``had I done B instead of A, would the outcome have been better?'', the kind of counterfactual reasoning that prediction-only world models cannot support, and the kind that safety increasingly demands as agents gain authority to act. Being able to predict the future state~\cite{assran2025v} may be insufficient: with counterfactual reasoning we enable prescriptive analysis, that is, the ability to answer \emph{How to?} questions and thus reasoning on how to reach the desired future state (which can also be optimized based on cost functions).

\paragraph{Models: causality-powered, multi-level training.}
The same causal structure that answers human and agent queries is also a powerful \emph{inductive bias} for the learning systems in the ecosystem. Conditioning training and fine-tuning on causal structure—rather than on raw correlations—suppresses spurious shortcuts, improves robustness under distribution shift, and reduces the number of samples needed to reach a given level of performance~\cite{kaddour2025causal,jiao2024causal}. Because the CWS exposes causal structure not as a single global object but as a hierarchy of local and global views, this causality-powered training can be applied \emph{at every level of the ecosystem}: a model local to one source can be trained against that source's local causal view, a subsystem model against an integrated regional view, and an ecosystem-wide model against the global causal structure. The same substrate that supports multi-level inference therefore also supports multi-level, sample-efficient training—each model learning with the granularity of causal knowledge appropriate to its scope.

\paragraph{Multimodal by construction.}
The ecosystem's causal signals are spread across structured tables, free text, time series, images, and video. These sources form the bottom layer of the architecture in Figure~\ref{fig:arch}, entering the CWS through the integration layer below. A CWS must therefore integrate \emph{multimodal causality}: a return spike (tabular), its explanation in tickets (text), and the product photo that misled buyers (image) are one causal story told in three modalities. The layer must align them onto shared causal variables.
The obstacle is that this knowledge is scattered, heterogeneous, and constantly changing. That is precisely the regime that data integration was invented for, so we build the CWS on its foundations.

\section{System Overview: A Mediated Causal Schema}
\label{sec:system}
Classical data integration offers a unified query interface over many sources through a mediated schema, related to the sources by either of two mappings~\cite{lenzerini2002data,halevy2001views}. 
Throughout, 
We use \emph{view} in its data-integration sense: an abstraction that re-presents heterogeneous sources through a mediated/global schema with a shared vocabulary~\cite{lenzerini2002data,halevy2001answering}.
For example, a view over a relational table selects and renames columns.
In each case the view distills the contents of one source into a fragment expressed in the ecosystem's common terms. We call a view \emph{local} when it speaks for a single source and \emph{global} when it denotes the integrated whole.
In \emph{Global-As-View} (GAV) the global schema is defined as views over the sources; querying is easy but the schema is brittle when sources change. In \emph{Local-As-View} (LAV) each source is defined as a view over the global schema; sources evolve independently but answering a query becomes the harder problem of answering queries using views~\cite{lenzerini2002data,halevy2001views}.

Our proposal is to lift this duality from data to \emph{causal structure}: sources publish local views, a \emph{mediator}, an architectural component in the data-integration sense, reconciles them into a global SCM, and causal queries are answered against the mediated schema by rewriting them over the views.
A defining commitment of the CWS is that this structure is \emph{white-box}: it is an explicit causal DAG over named ecosystem variables, not a latent representation. Every variable, edge, and estimation is inspectable and carries its provenance, so an answer is accompanied by the mechanisms and assumptions it rests on: making the CWS auditable and interpretable by construction, and giving humans and agents a structure they can read, contest, and trust rather than a black box they must take on faith.

\paragraph{Local views.}
CWS begins with many heterogeneous data sources, each exposed as a \emph{local view} (the Local views box in Figure~\ref{fig:arch}). Sources may be tables, logs, graphs, documents, images, audio, or application stores; each view specifies the variables, entities, timestamps, granularity, instrumentation, access constraints, provenance, and local assumptions of the evidence it provides. Structured sources may expose variables directly, while unstructured sources use learned encoders to map raw signals onto named ecosystem variables, allowing heterogeneous evidence to contribute without forcing every source into one global schema.

\paragraph{The global view and the mediator.}
The mediator turns these local views into a global view over ecosystem variables (the Global view in Figure~\ref{fig:arch}). This is where ordinary data-integration work becomes causal infrastructure. The mediator cleans noisy source data, transforms heterogeneous formats into comparable representations, aligns entities and variables across sources, and manages metadata, schema mappings, governance, and lineage. Variable alignment maps differently named or differently instrumented fields onto shared causal variables, while data modeling determines how entities, relations, events, and temporal observations are represented in the global view. The result is not a simple union of source records, but a mediated representation in which local evidence can be compared, joined, and traced back to its origin. In this sense, the global causal view plays the role of a mediated schema for causality: it defines the shared variables and relationships over which do($\cdot$) and counterfactual queries will eventually be answered~\cite{pearl2009causality}.

\paragraph{Causal World System.}
The CWM is the causal object; the CWS is the system that builds, maintains, and serves it. Once local views have been integrated into a global view, the CWS runs the causal machinery: causal discovery, provenance-aware edge arbitration, identifiability checking, causal inference, and counterfactual reasoning. Discovery methods such as PC~\cite{spirtes2000causation}, GES~\cite{chickering2002ges}, FCI~\cite{richardson2002ancestral}, and continuous-optimization approaches such as NOTEARS~\cite{zheng2018notears} can propose structural relations, while provenance determines how much trust to place in each edge: an experimental intervention or A/B test licenses stronger claims than an observational log. The system then tracks which effects are identifiable using ID/IDC and do-calculus~\cite{shpitser2008complete}, and when effects must be transferred across populations or instruments, it invokes data-fusion and transportability machinery~\cite{bareinboim2016fusion}. The CWS therefore needs storage, indexing, tuning, and query optimization to answer repeated interventional and counterfactual queries efficiently, support incremental maintenance as sources change, and revalidate only affected edges, estimands, and certifications rather than rebuilding the entire global graph.

\paragraph{Multi-level inference.}
Because views exist at several granularities, the \emph{same} CWS can be queried at several levels: a source-local view answers fine-grained mechanism questions, a subsystem view answers team-level diagnostics, and the global view answers ecosystem-level prescriptions. The formal device that makes this consistent is causal abstraction: coarse and fine views are $\tau$-abstractions of one another~\cite{beckers2019abstraction,rubenstein2017abstraction}, so a coarse query is answered by marginalizing over mechanisms it does not depend on, with a guarantee that the answer agrees with the fine-grained model. Practically, this lets the mediator pick the cheapest altitude that still identifies the effect, answering an operator's aggregate query against a cluster-DAG abstraction while reserving the full mechanism graph for an agent reasoning about a single action. Because the graph is explicit at every altitude, each answer is also an explanation: the mediator can return not only the effect estimate but the path of mechanisms that produced it and the assumptions under which it is identified, an audit trail that a latent world model cannot expose. Multi-level inference is what lets one substrate serve an operator, a team, and an agent simultaneously.

\paragraph{Multi-level, sample-efficient training.}
The view hierarchy is also a training scaffold. Causal structure is a powerful inductive bias: a model that respects the invariant mechanisms of the CWS needs far fewer samples to generalize than one fitting raw correlations~\cite{scholkopf2021crl,li2024causaldl}. Concretely, the explicit skeleton can be injected into learning as invariance constraints 
~\cite{arjovsky2019irm,peters2016icp}, as a structural prior on attention or message-passing, or as a generator of counterfactually augmented training data; and the causal lens sharpens data selection itself, prioritizing samples that are informative about contested mechanisms~\cite{humane2025influence}. Because these constraints are available across the view hierarchy, we train causality-informed models at multiple levels: local predictors are constrained by their LAVs, subsystem models by intermediate views, and ecosystem models by the GAV. The global causal skeleton is shared across these models as a common regularizer. Since this skeleton is explicit rather than latent, each inherited constraint is inspectable: one can identify which mechanisms a model was required to respect.
Certified mechanisms transfer across levels, letting downstream models reuse them and reduce sample and compute costs.


\section{Challenges and Future Work}
\label{sec:challenges}
Realizing the CWS is a research agenda for the whole ecosystem, not a single project.

\paragraph{C1. Causal discovery and integration at ecosystem scale.}
We must discover and align causal structure across hundreds of evolving sources, combining observational logs with the natural experiments organizations already run (A/B tests, staged rollouts, outages). The core question is a causal generalization of schema mapping: how to merge partial, possibly conflicting local causal views into a sound global one, tracking identifiability throughout~\cite{spirtes2000causation,bareinboim2016fusion}.

\paragraph{C2. View maintenance under drift.}
Ecosystems are non-stationary: mechanisms shift, instrumentation changes, and sources appear or vanish. The CWS therefore needs incremental maintenance: detecting stale edges, unreliable counterfactual estimates, and LAV sources that require re-certification, analogous to materialized-view maintenance over causal mechanisms~\cite{Soonbo2024view, Zhuge98view,han2025implementing,Gupta1993,pachera2025}.

\paragraph{C3. Multimodal causal alignment.}
Mapping latent encodings from JEPA-style~\cite{assran2023self} models onto shared causal variables, with calibrated uncertainty, is open. We need encoders that expose which causal variables they implicate, not just dense embeddings.

\paragraph{C4. Counterfactual reasoning as an agent primitive.}
Agents must query the ecosystem cheaply enough to consult it before every consequential action, and the interface must compose with planning. 
This makes counterfactual reasoning not only an agent-interface problem, but also a database-style query-optimization problem: the system should support cost-aware planning of causal queries, reuse of cached or materialized causal views, indexing over relevant interventions and contexts, and incremental or approximate evaluation when exact reasoning is too expensive.

\paragraph{C5. Identifiability, trust, and governance.}
A shared causal layer that drives decisions and autonomous actions becomes critical infrastructure. When is a queried effect actually identifiable from the available views? Who may write to the CWS, how are contested edges adjudicated, and how is a counterfactual that authorized an agent's action audited after the fact? These governance questions are inseparable from the technical ones.

\section{Conclusion}
As AI shifts from single models to whole ecosystems of predictors, language models, and agents, the binding constraint is no longer predictive accuracy but the ability to reason about consequences. We have argued that what the ecosystem lacks is not a larger model but a shared causal substrate: a Causal World System that overlays its architectures, systems, and processes with an explicit, white-box causal graph—built on the foundations of data integration and structural causal models, and queried by humans for analytical and prescriptive insight, by agents for counterfactual deliberation, and by learning systems for sample-efficient, multi-level training. Realizing it demands progress across causal discovery and fusion, view integration and maintenance, abstraction, and provenance, no single community's problem, and precisely the kind of grand challenge that the entire AI ecosystem, from learning and databases to systems and safety, is positioned to take up together.

\balance
\bibliographystyle{ACM-Reference-Format}

\bibliography{references}

\end{document}

%% file: macros.tex
\usepackage{soul}

%% file: references.bib
@inproceedings{nam2026cjepa,
  title     = {{Causal-JEPA}: Learning World Models through Object-Level Latent Masking},
  author    = {Nam, Heejeong and Le Lidec, Quentin and Maes, Lucas and LeCun, Yann and Balestriero, Randall},
  booktitle = {Proceedings of the 43rd International Conference on Machine Learning (ICML)},
  year      = {2026},
  note      = {arXiv:2602.11389}
}

@article{bardes2023mcjepa,
  title   = {{MC-JEPA}: A Joint-Embedding Predictive Architecture for Self-Supervised Learning of Motion and Content Features},
  author  = {Bardes, Adrien and Ponce, Jean and LeCun, Yann},
  journal = {arXiv preprint arXiv:2307.12698},
  year    = {2023}
}

@article{chen2025vljepa,
  title   = {{VL-JEPA}: Joint Embedding Predictive Architecture for Vision-Language},
  author  = {Chen, Delong and Shukor, Mustafa and Moutakanni, Theo and Chung, Willy and Kasarla, Tejaswi and Bang, Yejin and Bolourchi, Allen and LeCun, Yann and Fung, Pascale},
  journal = {arXiv preprint arXiv:2512.10942},
  year    = {2025}
}

@article{assran2023ijepa,
  title   = {Self-Supervised Learning from Images with a Joint-Embedding Predictive Architecture},
  author  = {Assran, Mahmoud and Duval, Quentin and Misra, Ishan and Bojanowski, Piotr and Vincent, Pascal and Rabbat, Michael and LeCun, Yann and Ballas, Nicolas},
  journal = {Proceedings of the IEEE/CVF Conference on Computer Vision and Pattern Recognition (CVPR)},
  year    = {2023}
}

@techreport{lecun2022path,
  title       = {A Path Towards Autonomous Machine Intelligence},
  author      = {LeCun, Yann},
  institution = {OpenReview},
  year        = {2022},
  note        = {Position paper, version 0.9.2}
}

@article{kaddour2022causalml,
  title   = {Causal Machine Learning: A Survey and Open Problems},
  author  = {Kaddour, Jean and Lynch, Aengus and Liu, Qi and Kusner, Matt J. and Silva, Ricardo},
  journal = {arXiv preprint arXiv:2206.15475},
  year    = {2022}
}

@article{li2024causaldl,
  title   = {Causal Inference Meets Deep Learning: A Comprehensive Survey},
  author  = {Li, Licheng and others},
  journal = {Research},
  volume  = {7},
  pages   = {0467},
  year    = {2024},
  doi     = {10.34133/research.0467}
}

@article{humane2025influence,
  title   = {Influence Functions for Efficient Data Selection in Reasoning},
  author  = {Humane, Prateek and Cudrano, Paolo and Kaplan, Daniel Z. and Matteucci, Matteo and Chakraborty, Supriyo and Rish, Irina},
  journal = {arXiv preprint arXiv:2510.06108},
  year    = {2025}
}

@book{pearl2009causality,
  title     = {Causality: Models, Reasoning, and Inference},
  author    = {Pearl, Judea},
  edition   = {2nd},
  publisher = {Cambridge University Press},
  year      = {2009}
}

@article{scholkopf2021crl,
  title   = {Toward Causal Representation Learning},
  author  = {Sch{\"o}lkopf, Bernhard and Locatello, Francesco and Bauer, Stefan and Ke, Nan Rosemary and Kalchbrenner, Nal and Goyal, Anirudh and Bengio, Yoshua},
  journal = {Proceedings of the IEEE},
  volume  = {109},
  number  = {5},
  pages   = {612--634},
  year    = {2021}
}

@inproceedings{lenzerini2002data,
  title     = {Data Integration: A Theoretical Perspective},
  author    = {Lenzerini, Maurizio},
  booktitle = {Proceedings of the 21st ACM SIGMOD-SIGACT-SIGART Symposium on Principles of Database Systems (PODS)},
  pages     = {233--246},
  year      = {2002}
}

@article{halevy2001views,
  title   = {Answering Queries Using Views: A Survey},
  author  = {Halevy, Alon Y.},
  journal = {The VLDB Journal},
  volume  = {10},
  number  = {4},
  pages   = {270--294},
  year    = {2001}
}

@article{bareinboim2016fusion,
  title   = {Causal Inference and the Data-Fusion Problem},
  author  = {Bareinboim, Elias and Pearl, Judea},
  journal = {Proceedings of the National Academy of Sciences (PNAS)},
  volume  = {113},
  number  = {27},
  pages   = {7345--7352},
  year    = {2016}
}

@book{spirtes2000causation,
  title     = {Causation, Prediction, and Search},
  author    = {Spirtes, Peter and Glymour, Clark and Scheines, Richard},
  edition   = {2nd},
  publisher = {MIT Press},
  year      = {2000}
}

@inproceedings{beckers2019abstraction,
  title     = {Abstracting Causal Models},
  author    = {Beckers, Sander and Halpern, Joseph Y.},
  booktitle = {Proceedings of the AAAI Conference on Artificial Intelligence},
  year      = {2019}
}

@article{pachera2025,
author = {Pachera, Amedeo and Palmiotto, Mattia and Bonifati, Angela and Mauri, Andrea},
title = {What If: Causal Analysis with Graph Databases},
year = {2025},
issue_date = {July 2025},
publisher = {VLDB Endowment},
volume = {18},
number = {11},
issn = {2150-8097},
url = {https://doi.org/10.14778/3749646.3749671},
doi = {10.14778/3749646.3749671},
journal = {Proc. VLDB Endow.},
month = jul,
pages = {4009–4016},
numpages = {8}
}

@article{jiao2024causal,
  title={Causal inference meets deep learning: A comprehensive survey},
  author={Jiao, Licheng and Wang, Yuhan and Liu, Xu and Li, Lingling and Liu, Fang and Ma, Wenping and Guo, Yuwei and Chen, Puhua and Yang, Shuyuan and Hou, Biao},
  journal={Research},
  volume={7},
  pages={0467},
  year={2024},
  publisher={AAAS}
}

@article{kaddour2025causal,
  title={Causal machine learning: A survey and open problems},
  author={Kaddour, Jean and Lynch, Aengus and Liu, Qi and Kusner, Matt J and Ricardo, Silva},
  journal={Foundations and Trends in Optimization},
  volume={9},
  number={1-2},
  pages={1--247},
  year={2025},
  publisher={Emerald Publishing Limited}
}

@article{chickering2002ges,
    author    = {David Maxwell Chickering},
    title     = {Optimal Structure Identification With Greedy Search},
    journal   = {Journal of Machine Learning Research},
    volume    = {3},
    pages     = {507--554},
    year      = {2002}
  }

@article{richardson2002ancestral,
    author    = {Thomas Richardson and Peter Spirtes},
    title     = {Ancestral Graph {Markov} Models},
    journal   = {The Annals of Statistics},
    volume    = {30},
    number    = {4},
    pages     = {962--1030},
    year      = {2002}
  }

@inproceedings{assran2023self,
  title={Self-supervised learning from images with a joint-embedding predictive architecture},
  author={Assran, Mahmoud and Duval, Quentin and Misra, Ishan and Bojanowski, Piotr and Vincent, Pascal and Rabbat, Michael and LeCun, Yann and Ballas, Nicolas},
  booktitle={Proceedings of the IEEE/CVF conference on computer vision and pattern recognition},
  pages={15619--15629},
  year={2023}
}

@inproceedings{zheng2018notears,
  title     = {DAGs with NO TEARS: Continuous Optimization for Structure Learning},
  author    = {Zheng, Xun and Aragam, Bryon and Ravikumar, Pradeep and Xing, Eric P.},
  booktitle = {Advances in Neural Information Processing Systems},
  volume    = {31},
  year      = {2018}
}

@article{shpitser2008complete,
  title   = {Complete Identification Methods for the Causal Hierarchy},
  author  = {Shpitser, Ilya and Pearl, Judea},
  journal = {Journal of Machine Learning Research},
  volume  = {9},
  pages   = {1941--1979},
  year    = {2008}
}

@inproceedings{rubenstein2017abstraction,
  title     = {Causal Consistency of Structural Equation Models},
  author    = {Rubenstein, Paul K. and Weichwald, Sebastian and Bongers, Stephan and Mooij, Joris M. and Janzing, Dominik and Grosse-Wentrup, Moritz and Sch{\"o}lkopf, Bernhard},
  booktitle = {Proceedings of the 33rd Conference on Uncertainty in Artificial Intelligence},
  year      = {2017}
}

@article{arjovsky2019irm,
  title   = {Invariant Risk Minimization},
  author  = {Arjovsky, Martin and Bottou, L{\'e}on and Gulrajani, Ishaan
             and Lopez-Paz, David},
  journal = {arXiv preprint arXiv:1907.02893},
  year    = {2019}
}

@article{peters2016icp,
  title   = {Causal Inference by Using Invariant Prediction: Identification and Confidence Intervals},
  author  = {Peters, Jonas and B{\"u}hlmann, Peter and Meinshausen, Nicolai},
  journal = {Journal of the Royal Statistical Society: Series B (Statistical Methodology)},
  volume  = {78},
  number  = {5},
  pages   = {947--1012},
  year    = {2016},
  doi     = {10.1111/rssb.12167}
}

@article{zeng2025DAGsumm,
  title={Causal DAG Summarization},
  author={Zeng, Anna and Cafarella, Michael and Kenig, Batya and Markakis, Markos and Youngmann, Brit and Salimi, Babak},
  journal={Proceedings of the VLDB Endowment},
  volume={18},
  number={6},
  pages={1933--1947},
  year={2025},
  publisher={VLDB Endowment}
}

@inproceedings{mansouri2024object,
  title={Object centric architectures enable efficient causal representation learning},
  author={Mansouri, Amin and Hartford, Jason and Zhang, Yan and Bengio, Yoshua},
  booktitle={International Conference on Learning Representations},
  volume={2024},
  pages={830--853},
  year={2024}
}

@inproceedings{kocaoglu2018causalgan,
  title={CausalGAN: Learning Causal Implicit Generative Models with Adversarial Training},
  author={Kocaoglu, Murat and Snyder, Christopher and Dimakis, Alexandros G and Vishwanath, Sriram},
  booktitle={International Conference on Learning Representations},
  year={2018}
}

@inproceedings{geiger2022NN,
  title={Inducing causal structure for interpretable neural networks},
  author={Geiger, Atticus and Wu, Zhengxuan and Lu, Hanson and Rozner, Josh and Kreiss, Elisa and Icard, Thomas and Goodman, Noah and Potts, Christopher},
  booktitle={International Conference on Machine Learning},
  pages={7324--7338},
  year={2022},
  organization={PMLR}
}

@inproceedings{yaramakala2005speculative,
  title={Speculative Markov blanket discovery for optimal feature selection},
  author={Yaramakala, Sandeep and Margaritis, Dimitris},
  booktitle={Fifth IEEE International Conference on Data Mining (ICDM'05)},
  pages={4--pp},
  year={2005},
  organization={IEEE}
}

@article{chen2024using,
  title={Using causal inference to avoid fallouts in data-driven parametric analysis: A case study in the architecture, engineering, and construction industry},
  author={Chen, Xia and Sun, Ruiji and Saluz, Ueli and Schiavon, Stefano and Geyer, Philipp},
  journal={Developments in the Built Environment},
  volume={17},
  pages={100296},
  year={2024},
  publisher={Elsevier}
}

@inproceedings{zhang2026causal,
  title={Causal-tune: mining causal factors from vision foundation models for domain generalized semantic segmentation},
  author={Zhang, Yin and Zhang, Yongqiang and Zheng, Yaoyue and Raducanu, Bogdan and Liu, Dan},
  booktitle={Proceedings of the AAAI Conference on Artificial Intelligence},
  volume={40},
  number={15},
  pages={12916--12924},
  year={2026}
}

@article{assran2025v,
  title={V-jepa 2: Self-supervised video models enable understanding, prediction and planning},
  author={Assran, Mido and Bardes, Adrien and Fan, David and Garrido, Quentin and Howes, Russell and Muckley, Matthew and Rizvi, Ammar and Roberts, Claire and Sinha, Koustuv and Zholus, Artem and others},
  journal={arXiv preprint arXiv:2506.09985},
  year={2025}
}

@article{Soonbo2024view,
author = {Han, Soonbo and Ives, Zachary G.},
title = {Implementation Strategies for Views over Property Graphs},
year = {2024},
issue_date = {June 2024},
publisher = {Association for Computing Machinery},
address = {New York, NY, USA},
volume = {2},
number = {3},
url = {https://doi.org/10.1145/3654949},
doi = {10.1145/3654949},
journal = {Proc. ACM Manag. Data},
month = may,
articleno = {146},
numpages = {26}
}

@INPROCEEDINGS{Zhuge98view,
  author={Zhuge, Y. and Garcia-Molina, H.},
  booktitle={Proceedings 14th International Conference on Data Engineering}, 
  title={Graph structured views and their incremental maintenance}, 
  year={1998},
  volume={},
  number={},
  pages={116-125},
  doi={10.1109/ICDE.1998.655767}}

@article{han2025implementing,
  title={Implementing Views for Property Graphs},
  author={Han, Soonbo and Ives, Zachary G.},
  journal={SIGMOD Record},
  volume={54},
  number={1},
  year={2025},
  pages={},
  publisher={ACM}
}

@article{Gupta1993,
author = {Gupta, Ashish and Mumick, Inderpal Singh and Subrahmanian, V. S.},
title = {Maintaining views incrementally},
year = {1993},
issue_date = {June 1, 1993},
publisher = {Association for Computing Machinery},
address = {New York, NY, USA},
volume = {22},
number = {2},
issn = {0163-5808},
url = {https://doi.org/10.1145/170036.170066},
doi = {10.1145/170036.170066},
journal = {SIGMOD Rec.},
month = jun,
pages = {157–166},
numpages = {10}
}

@article{halevy2001answering,
  author  = {Alon Y. Halevy},
  title   = {Answering Queries Using Views: A Survey},
  journal = {The VLDB Journal},
  volume  = {10},
  number  = {4},
  pages   = {270--294},
  year    = {2001},
  doi     = {10.1007/s007780100054}
}
